# Stable Nash equilibria of medium access games under symmetric, socially altruistic behavior

G. Kesidis
CS&E and EE Depts
Penn State University
kesidis@engr.psu.edu

Y. Jin
Sogang University
Seoul, Korea
youngmi@sogang.ac.kr

A.P. Azad and E. Altman
INRIA Sophia Antipolis
University of Avignon, France
{amar.azad,eitan.altman}@sophia.inria.fr

**Abstract**

We consider the effects of altruistic behavior on random medium access control (slotted ALOHA) for local area communication networks. For an idealized, synchronously iterative, two-player game with asymmetric player demands, we find a Lyapunov function governing the "better-response" Jacobi dynamics under purely altruistic behavior. Though the positions of the interior Nash equilibrium points do not change in the presence of altruistic behavior, the nature of their local asymptotic stability does. There is a region of partially altruistic behavior for which both interior Nash equilibrium points are locally asymptotically stable. Variations of these altruistic game frameworks are discussed considering power (instead of throughput) based costs and linear utility functions. Also, for a power control game with a single Nash equilibrium, we show how its stability changes as a function of the altruism parameter.

## I. Introduction

Game theoretic models for telecommunication systems have recently been surveyed in [4]. They are motivated by the need to model the very significant effects of end-user behavior. In the Internet, TCP congestion control, a protocol presuming cooperative end-user behavior, has been exploited by end-users and their client applications acting in selfish ways. As common in some wireless settings, *e.g.*, tactical mobile ad hoc networks (MANETs), network nodes may

The work was supported by NSF CISE grants 0524202 and 0915928 and by a Cisco Systems URP gift.

engage in cooperative/coalitional or simply altruistic behavior with respect to some of their peers. Altruistic action can be for the purpose of routing [7], medium access, *etc*. In [10], the behavior of permanent seeder peers in on-line BitTorrent swarms (characterized as altruistic) is asymmetric; clearly if one peer is completely selfish and the other completely altruistic, then the greedy peer stands to benefit. We consider a symmetric situation where the players have similar communication priorities and degrees of altruism. This said, the *demands* of the players are, however, generally assumed asymmetric in the following.

In distributed systems, altruistic actions can easily be shown to be counter-productive in the presence of

- limited information and/or observation errors,
- excessive communication overhead to (securely) convey more accurate network state information, and/or
- deliberate injection of false observation information by enemy actors.

In these cases, selfish behavior alone will give better performance. In coalitional games, the cost of cooperation is typically weighed against its benefits, *e.g.*, [5]. Extensive prior work has considered the effects of imperfect learning of the utilities of other players and the parameters of the environment in which they all interact, *e.g.*, [11], [26], [25].

In this paper, we assume perfectly correct and complete information about the player who is the target of altruistic behavior, and this at no cost. Despite this assumption, we show that it is possible that *negative* performance effects are possible under altruistic behavior. In particular, we will consider iterative games where utilities are a function of average throughput. The games are “quasi stationary” in that the (mixed or pure) strategies of players were based on observed players’ actions over time [23].

That Nash equilibria of iterative games are not necessarily asymptotically stable is well known, *e.g.*, [22], [1], [27]. In slotted-ALOHA medium access under congestion, each user (player) $i$ attempts to transmit with probability $p_i$ (*i.e.*, his/her control variable or strategy), so that the probability of successful transmission is $\gamma_i = p_i \prod_{j \neq i}(1 - p_j)$. In [12], [13], using a Lyapunov function for arbitrary $N \geq 2$ players, noncooperative two-player ALOHA was shown to have two different interior[1] Nash equilibria though only one was locally asymptotically stable, see also [15]. These local stability properties will change in the presence of altruistic behavior studied herein. In some partially altruistic cases, *both* interior Nash equilibria will be locally asymptotically stable. Local asymptotic stability properties are generally important for robust performance in the presence of modeling, estimation and quantization errors, *e.g.*, [6]. We similarly explore simple distributed power control medium access dynamics having a single feasible equilibrium.

[1] *i.e.*, not including the stable boundary deadlock equilibrium.

We focus on two-player games because: (a) the communication overhead required to accurately inform altruistic behavior will not scale to many users, (b) the players are assumed to have different utilities so one of them could be a model for an aggregation of users, and (c) the purely altruistic dynamics have a vector field (and Lyapunov function) expressible in closed form.

We also assume synchronous player action; the effects of asynchronous player action can be considered using, *e.g.*, the methods of [9], [8]. Even in these more general scenarios, potentially unstable Nash equilibria for synchronous two-person sub-game would remain a significant challenge for altruistic behavior.

This paper is organized as follows. In Section II, we give an overview of our game-theoretic framework. In Section III, we give the main results on the stability properties of the Nash equilibria, as a function of the "degree of altruism", of a two-player, synchronously iterative ALOHA medium access game. In Section IV, we consider variations of our game framework in which the utility is linear (instead of strictly concave) and costs are based on energy expenditure (rather than throughput). In Section V, we study medium access by power control involving a single feasible Nash equilibrium point. Finally, we conclude with a brief summary.

## II. Altruistic games

By using his/her control action (strategy) $q_i$, the $i^{\text{th}}$ player seeks to maximize the composite of the *net* utilities $V_j(\gamma_j) := U_j(\gamma_j) - \gamma_j M$, *i.e.*,

$$\sum_j \alpha_{ij}(U_j(\gamma_j) - \gamma_j M), \tag{1}$$

where $i$ and $j$ index the players,

$$\sum_k \alpha_{ik} = 1 \quad \text{and} \quad \alpha_{ij} \geq 0 \ \ \forall i, j, \tag{2}$$

$\gamma M$ is the usage-based charge for service $\gamma$, and all utilities $U_i$ are strictly concave, *i.e.*,

$$U_i''(\gamma_i) \ < \ 0 \ \ \forall \gamma_i \in \mathbb{R}^+. \tag{3}$$

The player actions $q$ are related to the service $\gamma$ through the network's dynamics. For the example of slotted ALOHA used below,

$$\gamma_i \ = \ q_i \prod_{j \neq i} (1 - q_j),$$

where the (re)transmission probabilities $0 \leq q_i \leq 1$ for all $i$. So, the users need to be aware of each others' actions and utilities[2]. For all collective-action vectors $\underline{q}$, we will typically assume

$$\frac{\partial \gamma_i}{\partial q_i}(\underline{q}) > 0 \ \ \forall \underline{q}, i, \ \ \text{and} \ \ \frac{\partial \gamma_i}{\partial q_j}(\underline{q}) < 0 \ \ \forall \underline{q}, i \neq j. \tag{4}$$

### *A. Purely selfish/non-cooperative games*

Regarding player $i$'s net utility,

$$\frac{\partial V_i}{\partial q_i}(\underline{q}) \ = \ (U_i'(\gamma_i(\underline{q})) - M) \frac{\partial \gamma_i}{\partial q_i}(\underline{q}),$$

where we have now explicitly written the $\gamma_i$ as functions of $\underline{q}$. So, at a Nash equilibrium point (NEP) $\underline{\hat{q}}$ of a *non*-altruistic game, *i.e.*, where $\alpha_{ii} = 1$ for all $i$,

$$\gamma_i(\underline{\hat{q}}) \ = \ (U_i')^{-1}(M) \ =: \ y_i \ \ \forall i. \tag{5}$$

Note that by the concavity assumption of the $U_i$, *at* a NEP $\underline{\hat{q}}$ is

$$\frac{\partial^2 V_i}{\partial q_i^2}(\underline{\hat{q}}) \ = \ U_i''(\gamma_i(\underline{\hat{q}})) \left( \frac{\partial \gamma_i}{\partial q_i}(\underline{\hat{q}}) \right)^2 \ < \ 0.$$

That is, (5) and (3) are the conditions for $\underline{q}$ to be an NEP.

[2] In the purely selfish/non-cooperative games ($\alpha_{ii} = 1 \ \forall i$) of [14], actions were taken based on observations in quasi steady-state thus not requiring any coordination between presumed selfish/non-cooperative peers.

### B. Altruistic invariance of NEPs

*Claim 1:* Under (3), the NEPs of the purely selfish game ($\alpha_{ii} = 1$ for all $i$) are also NEPs under the altruistic objective (1) for all $[\alpha_{ij}]$ satisfying (2).

*Proof:* Note that for objectives (1), the first order condition for player $i$ is

$$\sum_j \alpha_{ij}(U_j'(\gamma_j(\underline{q})) - M)\frac{\partial \gamma_j}{\partial q_i}(\underline{q}) \quad = \quad 0.$$

So, clearly this condition is satisfied when $\underline{q} = \hat{\underline{q}}$, *i.e.*, under (5). Moreover, the second order condition *at* such a $\hat{\underline{q}}$ (again, under (5)) reduces to

$$\sum_j \alpha_{ij} U_j''(\gamma_j(\hat{\underline{q}})) \left(\frac{\partial \gamma_j}{\partial q_j}(\hat{\underline{q}})\right)^2 \quad < \quad 0.$$

□

When the $\gamma_j$ are all linear functions of $\underline{q}$, a *qualified* converse of the previous claim, *i.e.*, that there are no additional interior NEPs under altruistic behavior, is given in [18].

### C. Asymptotic stability of NEPs under symmetric altruism

Again, it is possible that NEPs are not asymptotically stable. Though the positions of the NEPs may not change under altruistic behavior, the nature of their stability may change with the $(\alpha_{ij})$ parameters (as in the bifurcation theory of dynamical systems). Exploring this issue for specific examples is the main goal of the balance of this paper.

In the following, we restrict our attention to "symmetric" altruism where for some fixed $\alpha \in [0, 1]$, $\alpha_{ii} = \alpha$ and $\alpha_{ij} = (1-\alpha)/(N-1)$ for all $N \geq 2$ players $i$, and all $j \neq i$. Symmetric altruism is consistent with players whose communication is of equal priority, though they may value their own communication differently through different utility functions.

## III. Two-player ALOHA game with symmetric altruism

In the following, we simplify to the symmetric case by assuming the $\alpha_{ii}$ are the same ($\alpha$) for all players $i$. For the selfish, non-cooperative ALOHA game (*i.e.*, $\alpha = 1$), the (re)transmission probabilities $q_i$ obey the following continuous-time Jacobi approximation (*e.g.*, (10) of [13]):

$$\dot{\underline{q}}(t) \quad = \quad \underline{F}(\underline{q}(t)) - \underline{q}(t), \tag{6}$$

where for players $i \in \{1, 2\}$:

$$F_i(\underline{q}) \;=\; \frac{y_i}{1 - q_{3-i}}.$$

The deadlock equilibrium $q_i = 1\ \forall i$ is stable in the purely noncooperative case, and opt-out equilibrium $q_i = 0\ \forall i$ is stable in the purely altruistic case; at either equilibrium point, all players have zero throughput. To address this problem, we can restrict $q_i \in [q_{\min}, q_{\max}]$ for all $i$, so that

$$F_i(\underline{q}) \;=\; \min\{q_{\max}, \max\{q_{\min}, \frac{y_i}{1 - q_{3-i}}\}\},$$

where large $q_{\max} < 1$ is set to prevent a deadlock (or "capture" by just one player) and small $q_{\min} > 0$ is set to prevent opt-out. Our focus in this section will be on the stability of interior Nash equilibria for concave utilities.

In a distributed system, the Jacobi approximation is further justified when players take small steps toward their currently optimal play (*i.e.*, only "better response" based on their current knowledge of the state of the game including the actions of the other players in particular, *i.e.*, fictitious play [23]). One reason for this is that players act simultaneously so the optimal plays may change significantly. Thus, small steps may avoid large oscillations in the network dynamics, *i.e.*, oscillations that will be harmful to performance (*e.g.*, the balance between tentative and aggressive action taken by TCP in its distributed congestion control strategy). Also, small steps may ensure convergence to stable interior equilibria, even under asynchronous updates, and avoid deadlock or opt-out boundary equilibria.

A Lyapunov function governing the Jacobi iteration (6) is a function $\Lambda$ such that for all times $t$, the inner product $\langle \nabla\Lambda(\underline{q}(t)), \dot{\underline{q}}(t)\rangle \le 0$. For the non-cooperative $N$-player game [12], [13],

$$\Lambda(\underline{q}) \;=\; -\prod_{i=1}^{N} \frac{y_i}{1 - q_i} + \sum_{i=1}^{N} \left( \frac{q_i}{1 - q_i} + \log(1 - q_i) \right) \prod_{j \neq i} y_j.$$

In the following, we will assume $N = 2$ players.

### *A. Purely altruistic game ($\alpha = 0$)*

The throughput of player $3 - i$ is $(1 - q_i)q_{3-i}$, so that the purely altruistic choice for player $i$ is $q_i = 1 - y_{3-i}/q_{3-i}$. Thus, player $i \in \{1, 2\}$ updates

$$q_i \;=\; 1 - \frac{y_{3-i}}{q_{3-i}} \;=:\; G_i(\underline{q}). \tag{7}$$

That is, replace $F$ with $G$ in (6).

By Claim 1, the fixed points for this purely altruistic game ($\alpha = 0$) include those of the purely selfish game ($\alpha = 1$), *i.e.*, both given by the solution of $y_i = q_i(1 - q_{3-i})$ for $i \in \{1, 2\}$.

*Claim 2:* The Lyapunov function of the purely altruistic game is

$$\Lambda^*(\underline{q}) \;\; = \;\; -\prod_{i=1}^{2}(1 - \frac{y_i}{q_i}) + \sum_{i=1}^{2} y_i \log q_i. \tag{8}$$

*Proof:*

$$\frac{\mathsf{d}\Lambda^*}{\mathsf{d}t}(\underline{q}) = \langle \nabla\Lambda^*(\underline{q}), \underline{\dot{q}} \rangle = -\sum_{i=1}^{2}(G_i - q_i)^2 \frac{y_i}{q_i^2} \leq 0.$$

□

### B. Stability of purely altruistic or purely cooperative behavior

Let $H^*$, respectively $H$, be the Hessian corresponding to $\Lambda^*$, respectively $\Lambda$, *i.e.*, $H^*_{ij} = \partial^2\Lambda^*/\partial q_i \partial q_j$ for $i, j \in \{1, 2\}$.

*Claim 3:* (a) The NEP $\underline{q}$ is locally stable under purely altruistic ($\alpha = 0$) behavior

$$\Leftrightarrow \;\; \sigma^* \; := \; \frac{y_1 y_2}{q_1^2 q_2^2} \;\; < \;\; 1. \tag{9}$$

(b) The NEP $\underline{q}$ is locally stable under purely selfish ($\alpha = 1$) behavior

$$\Leftrightarrow \;\; \sigma \; := \; \frac{y_1 y_2}{(1 - q_1)^2(1 - q_2)^2} \;\; < \;\; 1. \tag{10}$$

*Proof:* $H^*$ is positive definite (*i.e.*, with two positive, real eigenvalues) at the NEP $\underline{q}$ (*i.e.*, where $\underline{G}(\underline{q}) = \underline{q}$) if and only if (9). Similarly, $H$ is positive definite at the NEP $\underline{q}$ (*i.e.*, where $\underline{F}(\underline{q}) = \underline{q}$) if and only if (10). □

The example selfish two-player ALOHA game of [12], [13] with $(y_1, y_2) = (8/15, 1/15)$ had NEPs $(q_1, q_2) = (2/3, 1/5)$ and $(4/5, 1/3)$:

| NEP $\underline{q}$ | $\sigma$ | $\sigma^*$ |
|---|---|---|
| $(2/3, 1/5)$ | $1/2$ | $2$ |
| $(4/5, 1/3)$ | $2$ | $1/2$ |

So, for purely altruistic (cooperative $\alpha = 0$) actions, (4/5,1/3) is the locally asymptotically stable NEP but (2/3,1/5) is unstable. On the other hand, for purely selfish (non-cooperative $\alpha = 1$) actions, (2/3,1/5) is the locally stable NEP but (4/5,1/3) is unstable. Clearly, $\sigma < 1$ and $\sigma^* < 1$ is not possible if $q_1 + q_2 < 1$, but note that $q_1 + q_2 > 1$ for the NEP $(4/5, 1/3)$.

### C. Numerical example for partial altruism

Given the current play $q_{3-i}$ of player $3-i$, let $Q_i(\alpha, q_{3-i})$ be the play of the $i^{\text{th}}$ player that maximizes (1), *i.e.*, $Q$ is used instead of $F$ in (6). In particular,

$$Q_i(\alpha, q_{3-i}) = \begin{cases} F_i & \text{if } \alpha = 1 \\ G_i & \text{if } \alpha = 0 \end{cases}$$

with no such closed-form expression available for $Q_i$ when $0 < \alpha < 1$.

For utilities of the form

$$U_i(\gamma) \;\; = \;\; M(1+y_i^2)\arctan(\gamma), \tag{11}$$

we numerically evaluated the local asymptotic stability of the two NEPs. That is, we chose utilities normalized by the price $M$ so that we can simply take $M = 1$. Stability can be ascertained by the Hartman-Grobman theorem (linearizing the dynamics at the equilibrium point and checking the eigenvalues of the Jacobian), *i.e.*, even if a Lyapunov function is not available.

Note from the following table that the stability of the NEP (4/5,1/3) changes at $\alpha \approx 0.58$ and similarly the stability of the NEP (2/3,1/5) changes at $\alpha \approx 0.42$.

| NEP$\backslash\alpha \in$ | $[0, 0.42)$ | $[0.42, 0.58]$ | $(0.58, 1]$ |
|---|---|---|---|
| $(2/3, 1/5)$ | unstable | stable | stable |
| $(4/5, 1/3)$ | stable | stable | unstable |

So, in the range $0.42 \leq \alpha \leq 0.58$ both interior NEPs are locally asymptotically stable. The contours of the Lyapunov function $\Lambda$ are given in Figure 1(a) and of $\Lambda^*$ in Figure 1(c), where one clearly sees the saddle contour in the latter. Given the large interval about $\alpha = 0.5$ where both NEP are stable, by continuity we expect that the this condition will hold for all sufficiently small asymmetries in altruism between the players, *i.e.*, $\alpha_{11} \neq \alpha_{22}$.

### D. Discussion: Alternative altruistic strategies

We now give a Lyapunov function *as a function of* $\alpha$ for an alternative altruistic play, *i.e.*, one that does not necessarily maximize (1).

Generally, $Q_i(\alpha, q_{3-i})$ will be a non-linear function of $\alpha$. We can consider an alternative play that is a linear combination:

$$Q_i^o \;\; := \;\; \alpha F_i + (1-\alpha)G_i.$$

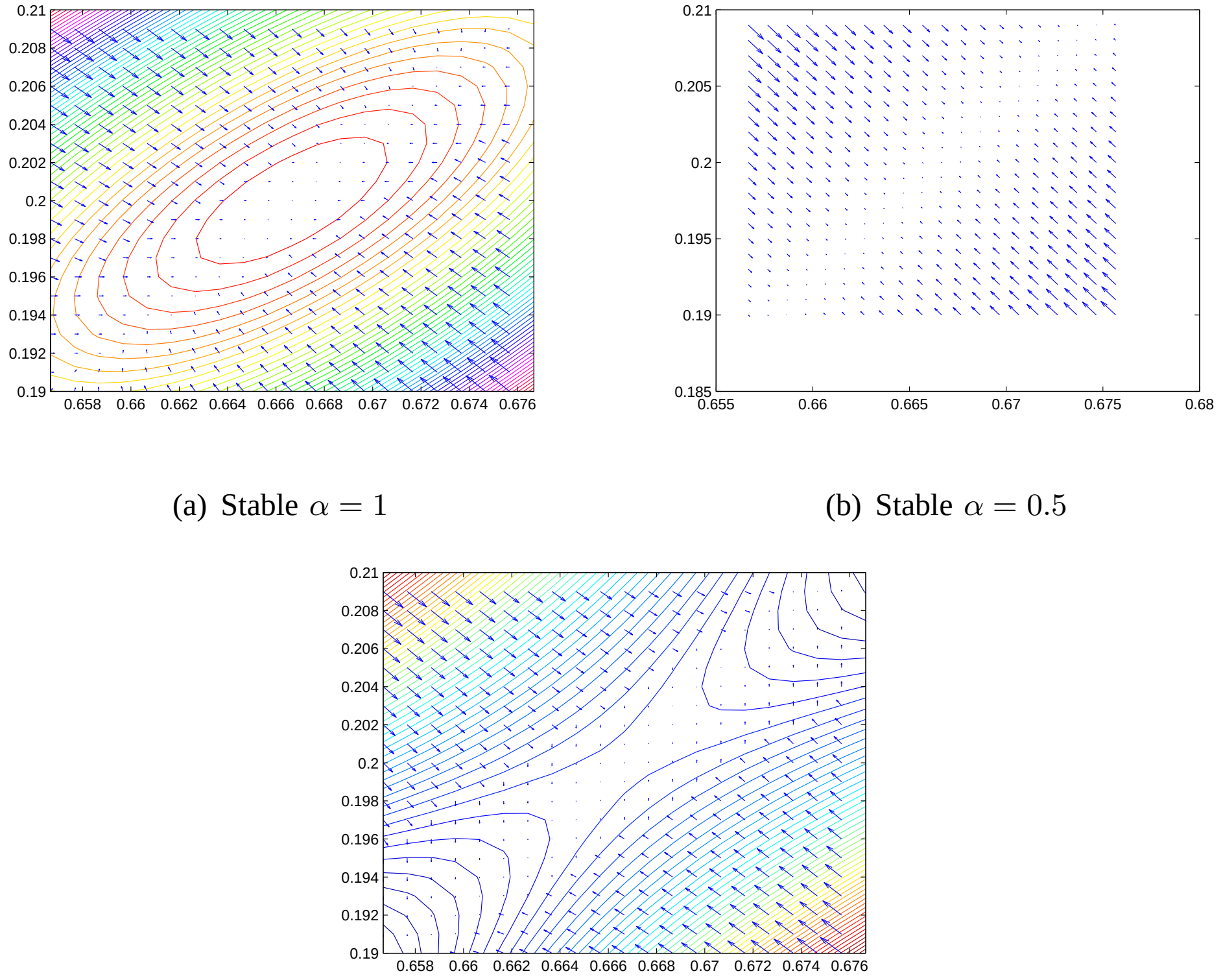

(a) Stable $\alpha = 1$

(b) Stable $\alpha = 0.5$

(c) Unstable (saddle) $\alpha = 0$

Fig. 1. The ALOHA NEP (2/3,1/5)

Moreover, instead of the purely altruistic $G_i$, consider

$$\tilde{G}_i = \left(1 - \frac{y_{3-i}}{q_{3-i}}\right) y_i^2 (q_i^{-1} - 1)^2,$$

corresponding to the $\alpha$-linear action

$$\tilde{Q}_i = \alpha F_i + (1-\alpha)\tilde{G}_i.$$

At a fixed point $\underline{q}$ for a purely altruistic game using $\tilde{G}_i$, $q_i$ can be easily shown to be an increasing function of $G_i y_i^2$.

*Claim 4:* The Jacobi iteration (6) using $\tilde{Q}$ instead of $F$ has the following Lyapunov function:

$$\Lambda_\alpha(\underline{q}) = -\alpha \prod_{i=1}^{2} \frac{y_i}{1-q_i} - (1-\alpha) \prod_{i=1}^{2} h_i(q_i) + \sum_{i=1}^{2} \left( \frac{q_i}{1-q_i} + \log(1-q_i) \right) y_{3-i}$$

where $h_i(q_i) := y_i - \; y_i^2/q_i$.

Note: We have modified the Lyapunov function (7) of [12], [13] by adding the $\alpha$ factors and the second term.

*Proof:*

$$\frac{\mathsf{d}\Lambda_\alpha}{\mathsf{d}t}(\underline{q}) =< \nabla\Lambda_\alpha(\underline{q}), \dot{\underline{q}} >= -\sum_{i=1}^{2}(\dot{q}_i)^2 \frac{y_{3-i}}{(1-q_i)^2} \leq 0.$$

□

We can similarly modify the purely altruistic iteration and find a Lyapunov function modified from $\Lambda^*$. Note that for these dynamics, the positions of interior NEPs may change as a function of $\alpha$.

## IV. Discussion: Variations of the game framework

In this section we discuss other extensions and variations of the synchronous, two-player game framework of the previous section.

### A. Linear utilities

If $U_i(\gamma) \equiv u_i\gamma$ for a scalar $u_i > M$, then the net utility of player $i$ is simply $\gamma_i(u_i - M)$.

So, in the purely selfish ($\alpha = 1$) case, player $i$ will simply maximize $\gamma_i$. That is, the selfish strategy is simply $q_i = 1$ if $q_j < 1 \; \forall j \neq i$. Thus, any play $\underline{q}$ such that at least one player $i$ uses the "pure" strategy $q_i = 1$ is a NEP. These include the deadlocked NEPs where two or more players choose $q = 1$ so that $\gamma = 0$ for all players.

In the purely altruistic, two-player case, each player $i$ will simply maximize $\gamma_{3-i} = q_{3-i}(1-q_i)$ (of the other player) over $q_i$, *i.e.*, choose $q_i = 0$ if $q_{3-i} > 0$. So, any play $\underline{q}$ such that at least one player uses the pure (opt-out) strategy $q = 0$ is an NEP.

For partially altruistic behavior, (1) with $0 < \alpha < 1$ remains a linear form. For the two-player game at equilibrium, $q_i = 1$ if

$$\begin{aligned} & \alpha(1-q_{3-i})(u_i - M) - (1-\alpha)q_{3-i}(u_{3-i} - M) \; > \; 0 \\ \Leftrightarrow \;\; q_{3-i} \; < \; & \frac{\alpha(u_i - M)}{\alpha(u_i - M) + (1-\alpha)(u_{3-i} - M)} \; =: \phi_{3-i}(\alpha). \end{aligned}$$

So, there are two stable NEPs (0,1) and (1,0) and the *unstable* saddle NEP $\underline{\phi}(\alpha)$ between them. More precisely, the following table indicates convergence trend based on the starting point $\underline{q}$.

| staring $\underline{q} \in$ | NEP $\underline{q}$ |
|---|---|
| $[0, \phi_1(\alpha)) \times (\phi_2(\alpha), 1]$ | (0,1) |
| $(\phi_1(\alpha), 1] \times [0, \phi_2(\alpha))$ | (1,0) |
| $[0, \phi_1(\alpha)) \times [0, \phi_2(\alpha))$ | $\underline{\phi}(\alpha)$ |
| $(\phi_1(\alpha), 1] \times (\phi_2(\alpha), 1]$ | $\underline{\phi}(\alpha)$ |

Note that *both* (0,1) and (1,0) are NEPs for $\alpha = 0$ and $\alpha = 1$ as well.

### *B. Power based costs*

Now instead of a cost of the form $M\gamma$ (*i.e.*, what the network charges for actual throughput), consider a cost of the form $Mq$ (*i.e.*, the average cost for power experienced by the user). Power-based costs are important for the context communications relying on limited energy supply in order to account for energy expenditure for failed communication due to interference (as here) and/or noise.

Under power-based costs, the NEPs may change position as a function of the degree of altruistic behavior, $\alpha$, *i.e.*, Claim 1 does not necessarily hold, *cf.*, Section V-E.

*1) Strictly concave utilities:* Here, the purely selfish update rule is

$$F_i^{\#}(\underline{q}_{-i}) \;=\; \frac{1}{\prod_{j\neq i} 1-q_i}(U_i')^{-1}\left(\frac{M}{\prod_{j\neq i} 1-q_i}\right). \tag{12}$$

For scalar parameters $\beta_i, u_i > 0$, consider utilities of the form

$$U_i(\gamma) = \frac{Mu_i}{\beta_i}\arctan(\beta_i\gamma) \;\;\Rightarrow\;\; (U_i')^{-1}(z) = \frac{1}{\beta_i}\sqrt{\frac{Mu_i}{z} - 1}.$$

For two players, (12) becomes

$$F_i^{\#}(q_{3-i}) \;=\; \frac{1}{\beta_i(1-q_{3-i})}\sqrt{u_i(1-q_{3-i})-1} \;\approx\; \frac{y_i}{\sqrt{1-q_{3-i}}},$$

where here $y_i := \sqrt{u_i}/\beta_i$ and the approximation holds when

$$u_i(1-q_{3-i}) \;\;\gg\;\; 1. \tag{13}$$

Under (13), the two-player Lyapunov function for purely selfish behavior is

$$\Lambda^{\#}(\underline{q}) = -\prod_i \frac{y_i}{\sqrt{1-q_i}} + \sum_i(\sqrt{1-q_i} + \frac{1}{\sqrt{1-q_i}})\prod_{j\neq i} y_i.$$

Under purely altruistic behavior, we see that

$$\frac{\partial}{\partial q_i}(U_{3-i}(\gamma_{3-i}) - Mq_{3-i}) \;\; = \;\; -q_{3-i}U'_{3-i}(\gamma_{3-i}) \;\leq\; 0.$$

Thus, the only NEPs $\underline{q}$ are such that $q_i = 0$ for at least one player $i$, *i.e.*, the opt-out action.

Suppose each utility $U_i$ is strictly concave only for an interval $[0, \hat{\gamma}_i]$ and $U'_i(\gamma_i(\underline{q})) = 0$ when

$$\gamma_i(\underline{q}) \;\; > \;\; \hat{\gamma}_i, \quad \forall i \in \{1, 2\}, \tag{14}$$

*i.e.*, the utility "saturates" after $\hat{\gamma}_i < 1$. Clearly here there will be additional NEPs under purely altruistic behavior: the set of plays $\underline{q}$ which jointly satisfy (14), assuming this set has non-empty intersection with the feasibility region $[0, 1]^2$ for $(q_0, q_1)$.

*2) Linear utilities:* Here again take $U_i(\gamma) = u_i\gamma$ for a scalar $u_i$. For the partially altruistic two-player case, $q_i = 1$ if

$$\begin{aligned} &\alpha[(1 - q_{3-i})u_i - M] - (1 - \alpha)q_{3-i}u_{3-i}) \;\; > \;\; 0 \\ \Leftrightarrow \;\; &q_{3-i} \;\; < \;\; \frac{\alpha(u_i - M)}{\alpha u_i + (1 - \alpha)u_{3-i})} \;\; =: \psi^{\mathrm{M}}_{3-i}(\alpha). \end{aligned}$$

The situation here is as in Section IV-A, except that for $\alpha = 1$ there is an additional unstable interior NEP, $\underline{\psi}^{\mathrm{M}}(1)$.

For $\alpha > 0$ and $u_i = u_j =: u$ (identical players), we can derive a price $\hat{M}(\alpha)$ for purely selfish behavior that mirrors $\alpha$-partially altruistic behavior:

$$\underline{\psi}^{\mathrm{M}}(\alpha) = \underline{\psi}^{\hat{\mathrm{M}}(\alpha)}(1) \;\; \Rightarrow \;\; \hat{M}(\alpha) = u - \alpha(u - M).$$

## V. Distributed power control games

Game-theoretic models for power control have been extensively studied, *e.g.*, [21], [20], [2], [24], [16], including consideration of issues of robust convergence to equilibria, *e.g.*, [3]. In the following, we consider a game played by unidirectional *flows* between pairs of nodes, including the mesh networking case where all one-hop flows share a single (gateway) node, but that node does *not* act as a central authority (base station or, possibly, mesh access point), rather the system is distributed in its decision making regarding transmission powers.

We further assume that a transmission attempt occurs in every time-slot by every player.

The signal to interference and noise ratio (SINR):

$$\mathsf{SINR}_i(\underline{q}) \;\; := \;\; \frac{q_i h_{ii}}{N + \sum_{j \neq i} h_{ji} q_j},$$

where $N$ is the ambient noise power, the power $q_i \geq 0$ pertains to the transmitter of flow $i$, and $h_{ji}$ are the path gains between the transmitter of flow $j$ and the receiver of flow $i$. Indeed, $\mathsf{SINR}_i$ is the SINR at the receiver of flow $i$. See Figure 2 illustrating two flow with transmitters $T_k$ and receivers $R_k$, $k \in \{0, 1\}$.

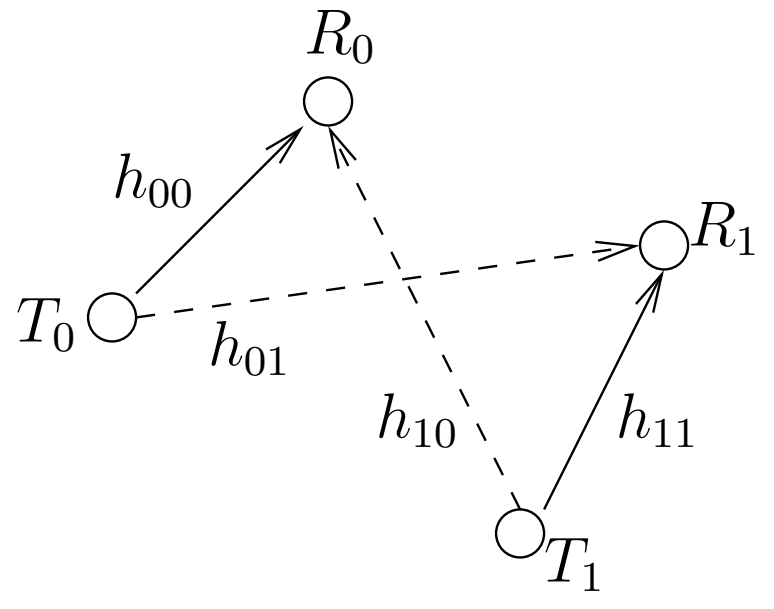

Fig. 2. Two interfering flows

The demands of each player $i$ $y_i := (U_i')^{-1}(M)$, as above.

Note that bidirectional links are not considered above due to self-interference at the transmitters/receivers. Typically, each way communication of a bidirectional link will be separated by TDMA, FDMA or CDMA/spread-spectrum means. In the TDMA setting for a distributed multihop wireless network, a spatial scheduling problem ensues, *e.g.*, [19].

### *A. From SINR to QoS, $\gamma$*

Shannon's expression for capacity $\log(1+\mathsf{SINR})$ is often used to map SINR to service quality. For different modulation frameworks, we can idealize

$$\gamma_i(\underline{q}) \;\; := \;\; \Gamma(\mathsf{SINR}_i(\underline{q}))$$

for correspondingly different increasing functions $\Gamma$, so that (4) holds [17]. If the number of bits per frame $n$ is large and $p_e$ an exponential function of SINR (as, *e.g.*, under DBPSK or GFSK modulation), then[3]

$$\Gamma(\mathsf{SINR}) \approx \exp(-n \exp(-\mathsf{SINR})).$$

We used this expression in our numerical studies below.

[3] *cf.*, the Appendix.

### B. The selfish game

In a quasi-stationary selfish game, the users observe their interference and user $i$ sets

$$q_i \;=\; \Upsilon_i(N + \sum_{j\neq i} q_j h_{ji}) \;=:\; F_i^o,$$

where

$$\Upsilon_i \;:=\; \Gamma^{-1}(y_i)/h_{ii}. \tag{15}$$

Note that this system is simply affine in $\underline{q}$ and a unique NEP $\underline{q}$ such that $\underline{q} = \underline{F}^o(\underline{q})$ can be determined if the matrix $I - \Psi$ is nonsingular where $\Psi_{ji} := h_{ji}\Upsilon_i \; \forall j \neq i$ and $\Psi_{ii} := 0 \; \forall i$ (again, here we are not considering constraints on power). That is, the NEP would be

$$\underline{q}^{\mathrm{T}} \;=\; N\underline{\Upsilon}^{\mathrm{T}}(I - \Psi)^{-1}. \tag{16}$$

For a two-player game, the Lyapunov function of the continuous-time Jacobi iteration $\dot{\underline{q}} = \underline{F}^o(\underline{q}) - \underline{q}$ is the quadratic form

$$\Lambda^o(\underline{q}) \;=\; \sum_i h_{i,3-i}\Upsilon_{3-i}(\frac{1}{2}q_i^2 - N\Upsilon_i q_i) - \prod_i q_i h_{i,3-i}\Upsilon_i.$$

The system is (globally) asymptotically stable with unique “interior” NEP if $I - \Psi$ has all eigenvalues with modulus $< 1$. Equivalently, we can specify stability in terms of the Hessians of the quadratic Lyapunov function, $\Lambda^o$. The result is that the NEP is stable if

$$\prod_i h_{i,3-i}\Upsilon_i \;<\; 1. \tag{17}$$

### C. The purely altruistic game

For an altruistic two-player game with information sharing as above, user $i$ sets

$$q_i \;=\; \frac{1}{h_{3-i,i}}(\frac{q_{3-i}}{\Upsilon_{3-i}} - N) \;=:\; G_i^o.$$

Note that $G_i^o$ is also a simple affine function. Again, we can show that the NEP (16) holds here too and similarly study its stability properties as in the non-cooperative case.

The Lyapunov function for the altruistic ($\dot{\underline{q}} = \underline{G}^o - \underline{q}$) two-player game is

$$\Lambda^+(\underline{q}) \;=\; \sum_i \frac{1}{h_{i,3-i}\Upsilon_i}(\frac{N}{h_{3-i,i}}q_i + \frac{1}{2}q_i^2) - \prod_i \frac{q_i}{h_{i,3-i}\Upsilon_i}.$$

The NEP stability condition for altruistic dynamics is

$$\prod_i h_{i,3-i}\Upsilon_i \quad > \quad 1, \tag{18}$$

which obviously cannot be true if (17) holds.

Recalling (15), we see a potentially beneficial role for symmetric altruism in the case where $\prod_i \Gamma^{-1}(y_i) > \prod_i h_{ii}/h_{i,3-i}$ so that (17) does not hold, and therefore (18) does.

### *D. Numerical example*

Consider an example where the frame sizes $n = 1024$ bits (128 bytes), the desired mean correct frame transmission probabilities $(y_0, y_1) = (.97, .98)$, the noise power $N = 1.0$ (so the transmission powers $(q_1, q_0)$ are normalized with respect to noise), the path gains[4] $h_{i,i} = 0.1$ and $h_{i,3-i} = 0.005$ for all $i$, and the utilities are of the arctan form (11).

The unique feasible NEP is $(q_0, q_1) = (224, 230)$ with corresponding SINRs of 10.4 and 10.8, respectively, and comparable noise and interference magnitudes. This NEP does not change position as the altruism parameter $\alpha$ changes, *i.e.*, Claim 1. The asymptotic stability condition for altruistic dynamics (18) does *not* hold for this example, giving a saddle contour for $\Lambda^+$ as in Figure 1(c).

### *E. Game framework variation: power-based costs*

With the cost $Mq$ instead of $M\gamma$, as $\alpha \downarrow 0$ (to purely altruistic behavior) the equilibrium point converged to an opt-out equilibrium, *i.e.*, $(q_0, q_1) = (0, 0)$, consistent with Section IV-B.

## VI. Summary

We considered synchronously iterative two-player medium access games under simple ALOHA or power-control dynamics. Assuming symmetric altruistic behavior (but not necessarily symmetric demand), we showed how the local asymptotic stability properties of the Nash equilibria changed as a function of the degree of altruism. Even assuming the necessary information is free of cost and perfect, such altruistic behavior may not have net beneficial effects in this regard.

[4]The coding strategy will, in many cases, additionally reduce the interference factor $h_{i,3-i}$ beyond propagation attenuation, *i.e.*, by a "processing gain" factor of the code, here assumed to be $13\text{dB} = 20 = h_{i,i}/h_{i,3-i}$.

In particular, partial altruistic behavior for an example ALOHA game can cause both interior, feasible Nash equilibria to be stable. A beneficial stabilizing effect is possible for some cases of a power control game when the NEP is unstable in the non-cooperative setting.

*Acknowledgement:* John F. Doherty for helpful discussions on power control.

## APPENDIX: FROM SINR TO QOS, $\gamma$

The following expressions for the bit error probability as a function of the modulation have been derived [17]:

$$p_e(\mathsf{SINR}) = \begin{cases} \frac{1}{2}\mathsf{erfc}(\sqrt{\kappa \cdot \mathsf{SINR}}) & \text{for GMSK} \\ \frac{1}{2}\exp\left(-\mathsf{SINR}\right) & \text{for DBPSK} \\ \frac{1}{2}\exp(-\frac{1}{2} \cdot \mathsf{SINR}) & \text{for GFSK} \\ \frac{1}{2}\mathsf{erfc}(\sqrt{\mathsf{SINR}}) & \text{for QPSK} \\ \frac{3}{8}\mathsf{erfc}(\sqrt{\frac{2}{5} \cdot \mathsf{SINR}}) & \text{for 16-QAM} \\ \frac{7}{32}\mathsf{erfc}(\sqrt{\frac{4}{21} \cdot \mathsf{SINR}}) & \text{for 64-QAM} \end{cases}$$

where $\kappa$ is a constant (that depends on the amount of redundancy in the coding and on the frequency band). In the absence of redundancy this gives the following expression for the prob-ability of correct reception of an $n$-bit packet assuming that the bit loss process is independent,

$$\Gamma(\mathsf{SINR}) = (1 - p_e(\mathsf{SINR}))^n.$$

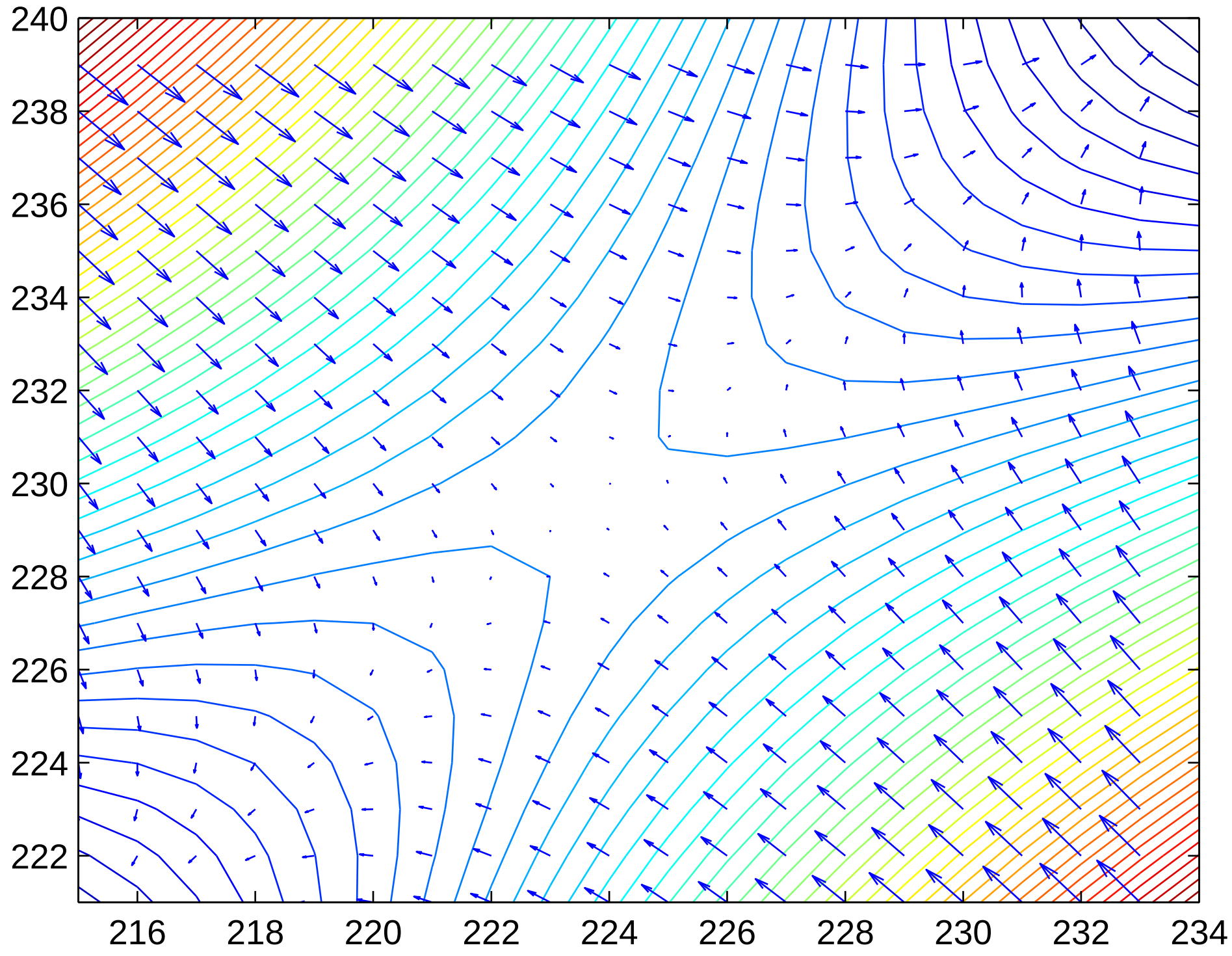

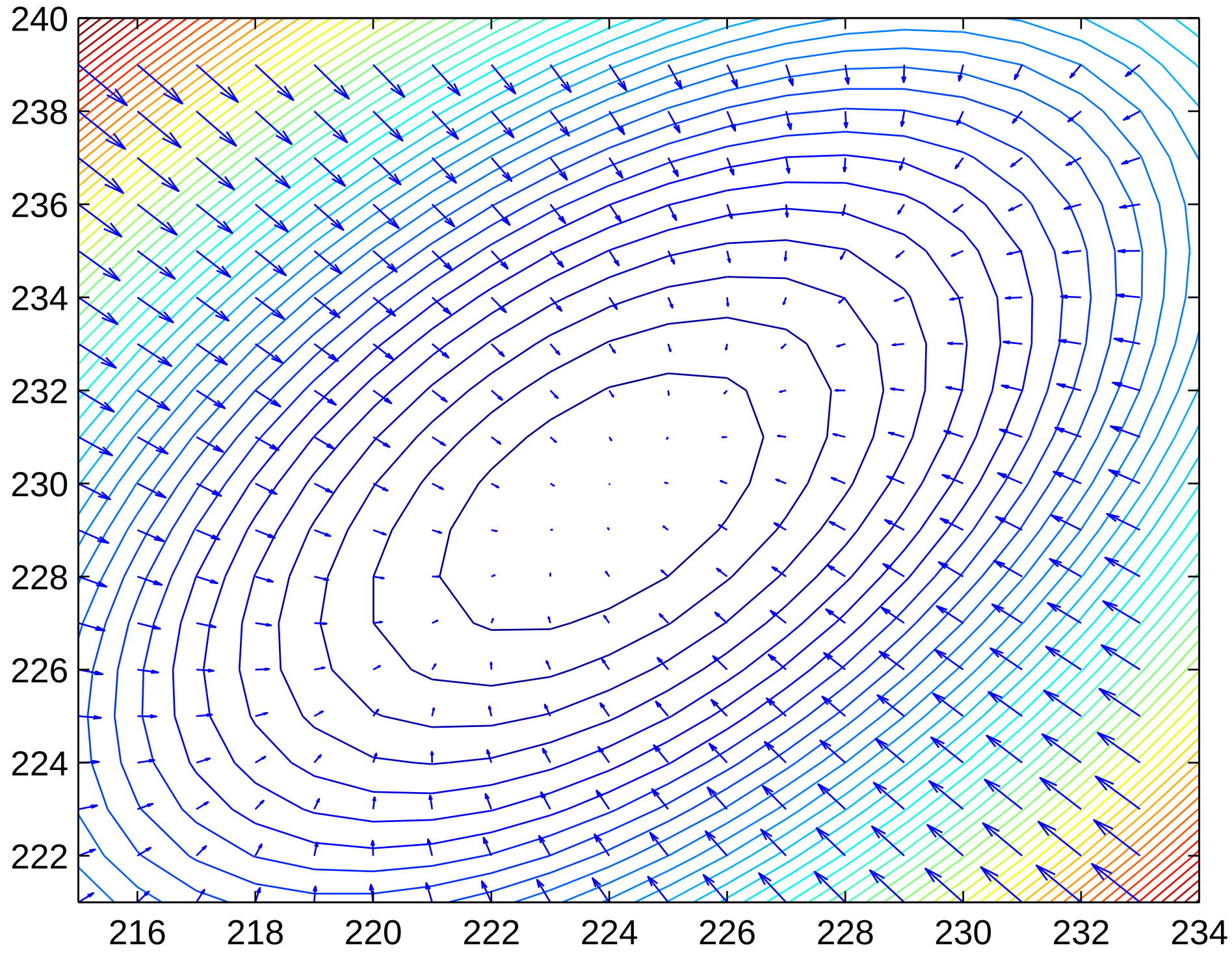

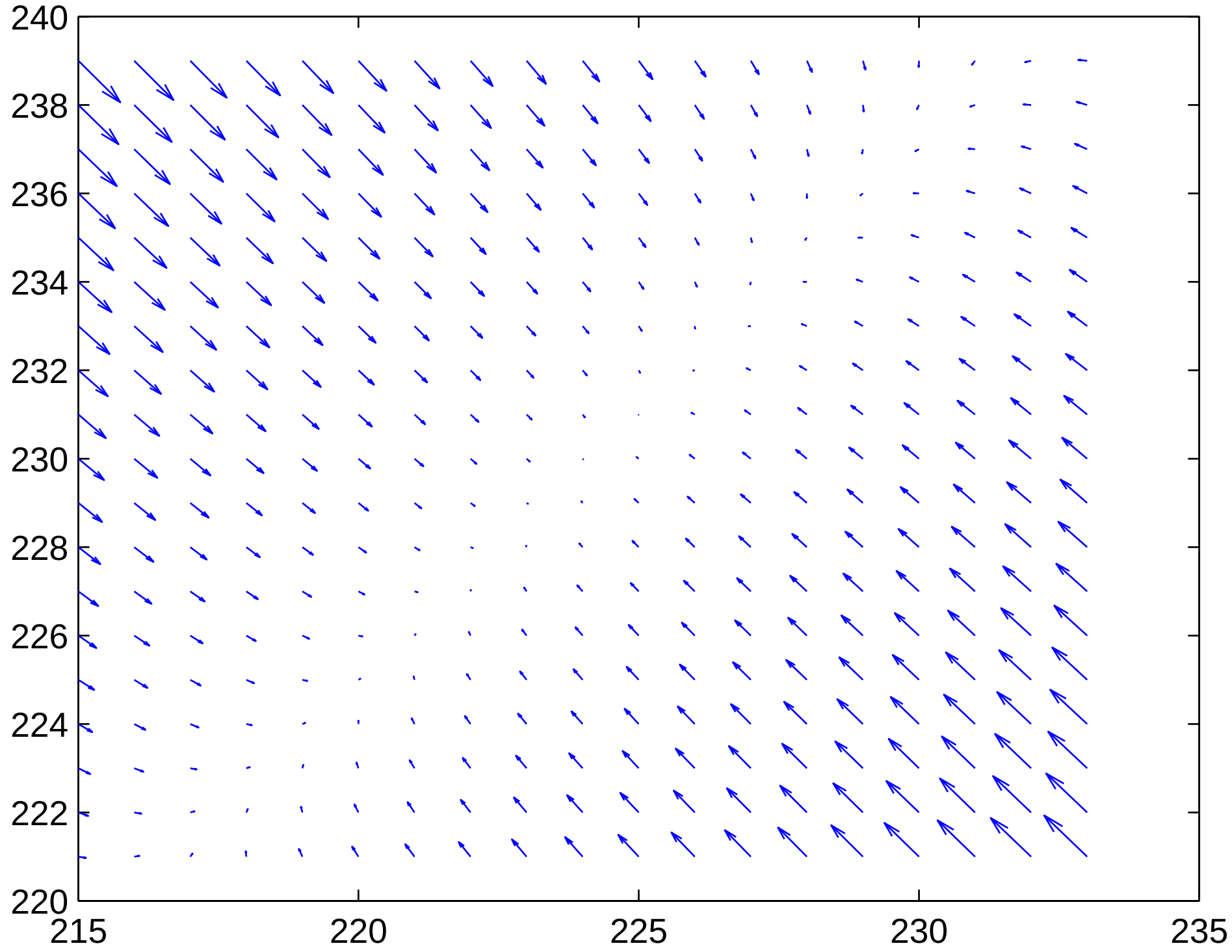

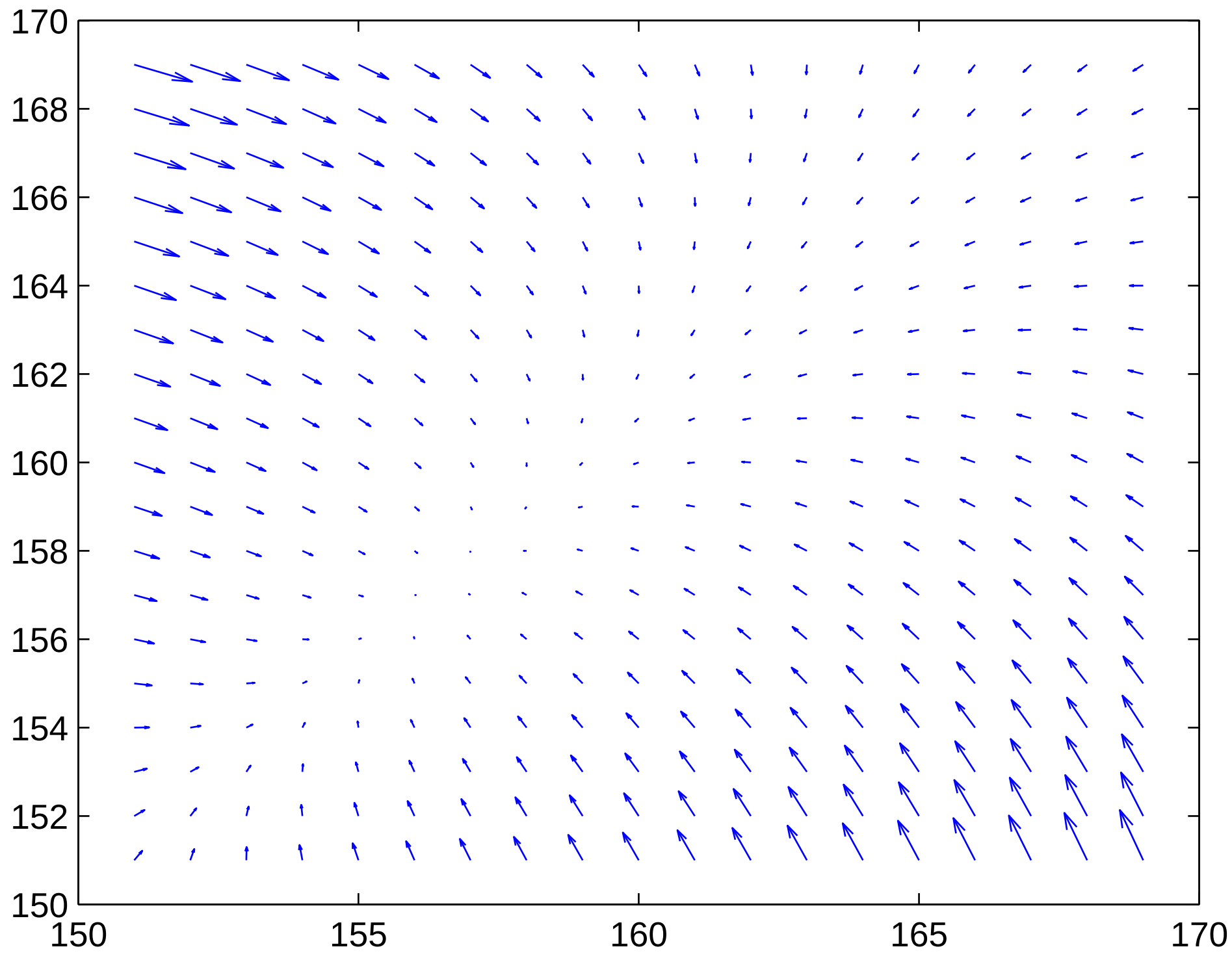

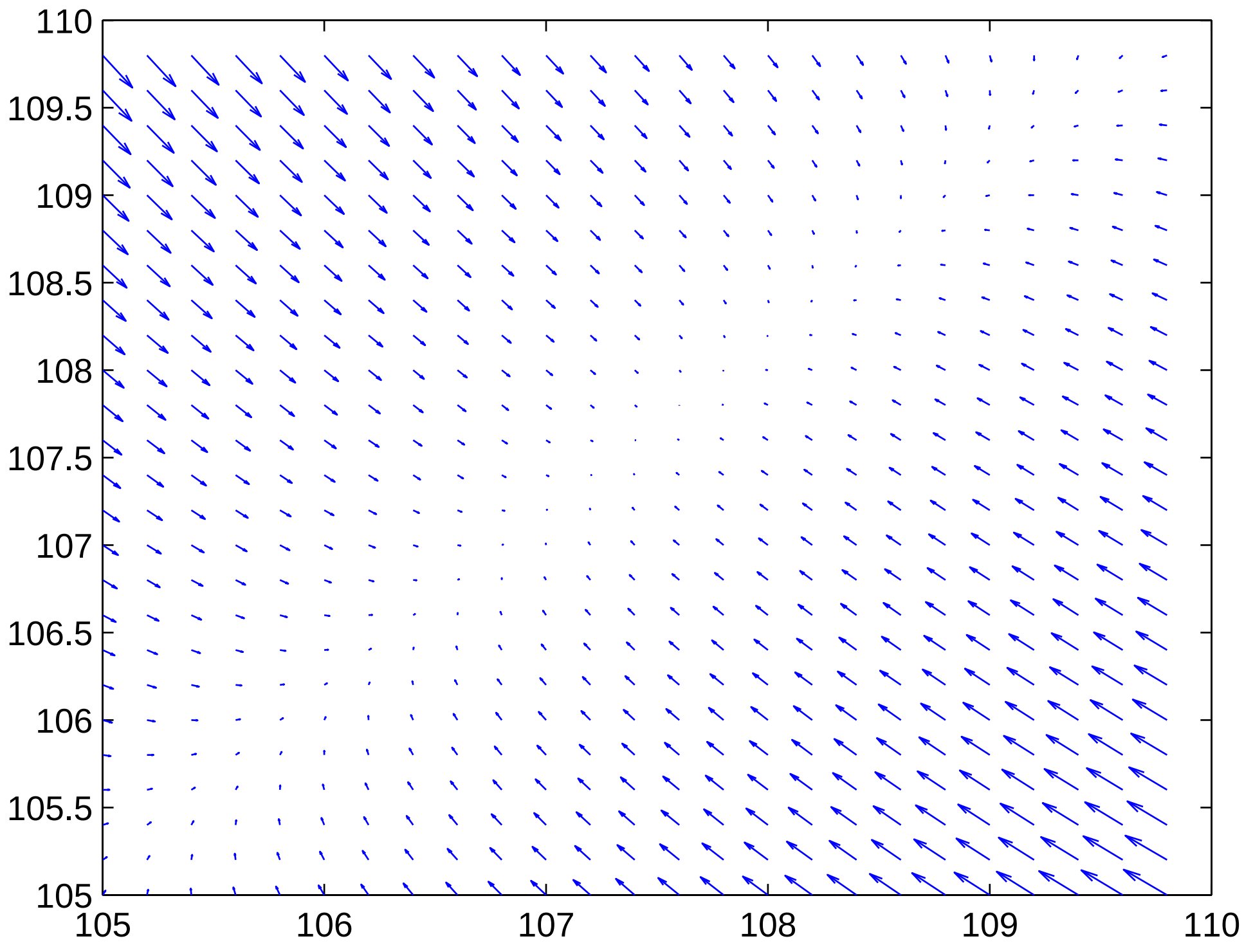